\documentclass[letterpaper, 10 pt, conference]{ieeeconf}  % Comment this line out
\IEEEoverridecommandlockouts                              % This command is only
\usepackage{cite}
\usepackage{amsmath,amssymb,amsfonts,bm}
\usepackage{bbm}
\usepackage{graphicx}
\usepackage{subfigure}
\usepackage{textcomp}
\usepackage{algorithm}
\usepackage{algpseudocode}

\usepackage{afterpage}
\usepackage{stfloats}   % 保证跨双栏公式可以放置在底部
\usepackage{cuted}

\usepackage{mathrsfs}
\usepackage{xcolor}
\allowdisplaybreaks[4]

\newtheorem{lemma}{Lemma}

\title{\LARGE \bf
Direct Adaptive Control of Grid-Connected Power Converters via Output-Feedback Data-Enabled Policy Optimization
}

\author{Feiran Zhao, Ruohan Leng, Linbin Huang, Huanhai Xin, Keyou You, Florian D\"{o}rfler % <-this % stops a space 
\thanks{F. Zhao is with the Department of Automation and BNRist, Tsinghua University, Beijing 100084, China, and the Department of Information Technology and Electrical Engineering, ETH Zurich, 8092 Zurich, Switzerland. (e-mail: zhaofe@control.ee.ethz.ch)} %
\thanks{R. Leng, L. Huang, and H. Xin are with the College of Electrical Engineering at Zhejiang University, Hangzhou 310027, China. (email: lengruohan@zju.edu.cn, hlinbin@zju.edu.cn, xinhh@zju.edu.cn)}
\thanks{K. You is with the Department of Automation and BNRist, Tsinghua University, Beijing 100084, China. (e-mail: youky@tsinghua.edu.cn.)} 
\thanks{F. D\"{o}rfler is with the Department of Information Technology and Electrical Engineering, ETH Zurich, 8092 Zurich, Switzerland. (e-mail: dorfler@ethz.ch)}
}

\begin{document}

\maketitle
\thispagestyle{empty}
\pagestyle{empty}

%%%%%%%%%%%%%%%%%%%%%%%%%%%%%%%%%%%%%%%%%%%%%%%%%%%%%%%%%%%%%%%%%%%%%%%%%%%%%%%%
\begin{abstract}
% The stabilization of power converter systems has been acknowledged as an important but challenging problem due to unknown nonlinear dynamics and sudden disturbances from external grid network. 
Power electronic converters are becoming the main components of modern power systems due to the increasing integration of renewable energy sources. However, power converters may become unstable when interacting with the complex and time-varying power grid. In this paper, we propose an adaptive data-driven control method to stabilize power converters by using only online input-output data. Our contributions are threefold. First, we reformulate the output-feedback control problem as a state-feedback linear quadratic regulator (LQR) problem with a controllable non-minimal state, which can be constructed from past input-output signals. Second, we propose a data-enabled policy optimization (DeePO) method for this  non-minimal realization to achieve  efficient output-feedback adaptive control. Third, we use high-fidelity simulations to verify that the output-feedback DeePO can effectively stabilize grid-connected power converters and quickly adapt to the changes in the power grid.
\end{abstract}

% Power electronic converters are becoming a main component of modern power systems due to the developments of renewable energy, high-voltage DC systems, electric vehicles, etc. However,  power converters may become unstable in practice when interacting with the power grid. One of the main reasons behind is that the power grid dynamics are variable and unknown from the perspective of a converter, and can hardly be considered in the converter's off-line control design. This problem can potentially be solved by leveraging the online data that captures the closed-loop dynamical interaction between the converter and the grid. To this end,
% in this paper, we propose a computationally efficient data-driven control method, called data-enabled policy optimization (DeePO), to stabilize power converters by using only input-output data. Firstly, we propose a covariance parameterization of partially observed linear systems with input-output data. Secondly, we develop a DeePO algorithm, which performs policy gradient computation in real time to achieve adaptive and optimal control using online closed-loop input-output data. Moreover, we showcase how DeePO can effectively stabilize power converter systems and quickly adapt to the changes in power system dynamics.

%%%%%%%%%%%%%%%%%%%%%%%%%%%%%%%%%%%%%%%%%%%%%%%%%%%%%%%%%%%%%%%%%%%%%%%%%%%%%%%%
\section{Introduction}\label{sec: intro}
Modern power systems feature a large-scale integration of power electronic converters, as they act as interfaces between the AC power grid and renewable energy sources, high-voltage DC (HVDC) systems, energy storage systems, and electric vehicles~\cite{milano2018foundations}. The large-scale integration of converters is fundamentally changing the power system dynamics, as they are significantly different from traditional synchronous generators (SGs). Usually, multiple nested control loops, based on fixed-parameter PI regulators, are needed in converters to achieve voltage, current, and power regulations. Under these control loops, power converters exhibit complicated interaction with the power grid and may easily tend to be unstable due to unforeseen grid conditions~\cite{huang2018an, huang2019grid, wang2017unified}. Such instability issues have been widely observed in practice~\cite{chen2020wind}, which poses challenges to the secure operation of modern power systems and impedes further integration of renewables. 

The instability in converter systems is caused by the closed-loop interaction between the converter and the complex power grid, which often occurs when the converter's control strategy does not fit into the grid characteristics~\cite{harnefors2007modeling, cespedes2013impedance}. Hence, the control design of converters should take into account the power grid dynamics for the sake of stability. However, the power grid is unknown, nonlinear, and time-varying from the perspective of a converter. Moreover, the grid structure and parameters are difficult to obtain in real time. Hence, it is nearly impossible to establish an exact dynamical model of a power grid for the control design of converters. As a remedy, engineers often use an overly simplified model for the power grid (e.g., an infinite bus) and tune the controller based on engineering experience and iterative trial-and-error approaches, which can be expensive, time-consuming, and lacking stability guarantees due to the model mismatch. While existing robust control methods can be used to handle the model mismatch~\cite{zhou2017large, weiss2004h}, they usually lead to conservative controllers when large changes appear in the power grid (e.g., tripping of transmission lines or even HVDC stations). Ideally, the controller of converters should be \textit{adaptive}, i.e., it is able to perceive and quickly adapt to changes in the power grid by using online data.

%However, the interactions among the multiple control loops of the power converters and the AC grid usually lead to instability issues of power systems across various time scales \cite{huang2018an, huang2019damping}. Thus, it is essential to ensure stability of power converters via control design.

%There are two main challenges in designing a stabilizing controller for power converters. First, an exact dynamical model of power systems is difficult to establish due to its nonlinear and high-order nature. Second, the power converter is usually in a grid network and hence can be affected by varying external grid conditions. Moreover, there exist disturbances at fast time scales, such as the intermittency of renewable energy. This means that the controller should be able to \textit{adapt} to sudden change of dynamics.

Recently, there has been a renewed interest in \textit{direct} data-driven control, which bypasses the system identification (SysID) step and learns the controller directly from a batch of persistently exciting data~\cite{zhao23global, zhao2024convergence,coulson2019data,dorfler21bridging,dorfler2021certainty,chiuso2023harnessing, de2019formulas, van2020data, liu2024learning, kang2023minimum}. This approach is end-to-end, easy to implement, and has seen many successful applications~\cite{markovsky2021behavioral,markovsky2023data,dorfler2023data}. Following this line, our previous work proposes a direct data-driven linear quadratic regulator (LQR) method~\cite{zhao2023data,zhao2024data}. It is adaptive in the sense that the control performance is improved in real time by using online closed-loop data. We call this method \textbf{D}ata-\textbf{e}nabl\textbf{e}d \textbf{P}olicy \textbf{O}ptimization (DeePO), where the policy is parameterized with sample covariance of input-state data and updated using gradient methods.
DeePO is computationally efficient, meets provable stability and convergence guarantees for linear time-invariant systems, and has successful real-world applications \cite{persson2025adaptive}. However, the DeePO method \cite{zhao2024data} works only on the state-feedback control problem, which is not the case for the power converter system.

In this paper, we propose an output-feedback DeePO method to mitigate oscillations in power converter systems. To this end, we
first reformulate the output-feedback control problem as a state-feedback LQR problem with a controllable non-minimal state, which can be constructed from input-output signals. This is achieved by the state reduction method in \cite{alsalti2023notes}. Then, we apply DeePO to this controllable non-minimal realization to achieve output-feedback adaptive control. Finally, we apply the output-feedback DeePO algorithm to stabilize power converter and direct-drive wind generator systems, both of which are unknown, state-unmeasurable, and encounter sudden change of dynamics due to grid changes. Simulation results show that output-feedback DeePO enables efficient online adaptation and effectively prevents instabilities in grid-connected power converters.
Compared with our previous works applying data-enabled predictive control (DeePC) to power converters \cite{markovsky2023data,deepc-appl}, DeePO offers a significantly reduced online computational burden, making it more suitable for scenarios where the processor cannot solve a quadratic program in real time.

The rest of the paper is organized as follows. Section \ref{sec:formu} formulates the stabilization problem of power converter systems. Section \ref{sec: output} proposes a controllable non-minimal realization of the output feedback system. Section \ref{sec:deepo} proposes an output-feedback DeePO algorithm for adaptive control. Section \ref{sec: sim} performs simulations on the converter systems. Conclusions are made in Section \ref{sec:conclu}.

\textbf{Notation.} We use $I_n$ to denote the $n$-by-$n$ identity matrix.  We use $\rho(\cdot)$ to denote the spectral radius of a square matrix. We use $A^\dagger$ to denote the pseudoinverse of a matrix $A$. We use $(S)_i$ to denote the $i$-th column of a block matrix $S$.

\section{Stabilization of grid-connected power converters}\label{sec:formu}

In this section, we introduce the stabilization problem of DC-AC power converter systems, taking into account both the DC-side and AC-side dynamics. Notice that power converters are widely used as the interface between the AC power grid and a DC source, such as lithium battery-based energy storage systems or direct-drive wind generators. The lithium batteries generate a constant DC voltage, and the converter aims to regulate the active and reactive power, as shown in Fig.~\ref{fig.1}. By comparison, when used as the grid interface of direct-drive wind generators, the converter needs to regulate the DC voltage and the reactive power, as will be shown in Section~V. In both cases, the converter plays an important role in maintaining stable and reliable power transfer between the DC side and the AC side. 

%Then and the corresponding challenges, including unknown model and time-varying dynamics. 

%These challenges form the motivations for proposing the data-driven DeePO algorithm in this paper.

\subsection{Power converter systems}

\begin{figure}[!] 
    \centering 
    \vspace{1.5mm}
    \includegraphics[width=0.98\linewidth]{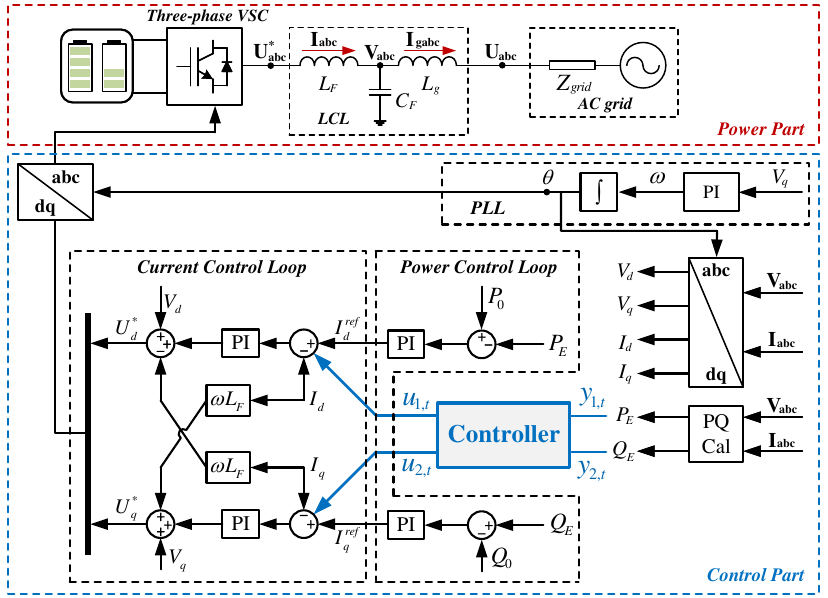} 
    \vspace{-3mm}
    \caption{One-line diagram of a grid-connected power converter. Here the DC side is connected to lithium batteries, while it can also be wind turbines.} 
    \label{fig.1} 
\end{figure}

Consider the grid-connected power converter system in Fig. \ref{fig.1}, which has a phase-locked loop (PLL), power/current control loops, and (abc/dq) coordinate transformation blocks \cite{harnefors2007modeling}. Due to proprietary manufacturer models and the complexity of the power grid, the power converter together with the grid is a black-box system for the subsequent stabilization control design. While inherently nonlinear, the system can be linearized around its equilibrium point for analysis. Note that the time-varying nature of the power grid may induce a time-varying operating point. Without loss of generality, consider the state-space model linearized at the origin
\begin{equation}\label{equ:outputsys}
\begin{aligned}
x_{t+1} &= A x_t + B u_t + d_t, \\
y_t &= C x_t + v_t,
\end{aligned}
\end{equation}
where $x_t \in \mathbb{R}^{n}$ is the state variable (could be unmeasurable, e.g., variables in the power grid side),
$u_t\in\mathbb{R}^{m}$ is the control input (e.g., additional signals added to the current references; see Fig.~\ref{fig.1}),
$y_t \in \mathbb{R}^{p}$ is the output (e.g., active and reactive power in our setting), $d_t$ is the process noise, and $v_t$ is the measurement noise. The unknown system $(A, B, C)$ is controllable and observable, but it may be subject to sudden impedance changes due to external grid changes, e.g., changes of $Z_{\rm grid}$ in Fig.~\ref{fig.1}. Such events may lead to poorly damped or even unstable oscillations.

%Note that the nonlinearity of the system enters into \eqref{equ:outputsys} via the disturbance and measurement noise $w_t,v_t$.

\afterpage{
	\begin{figure*}[b]
		\normalsize
		\hrulefill 
		\begin{equation}\label{equ:nonnimimal_realization}
			\begin{aligned}
				\xi_{t+1} &= \begin{bmatrix}
					0 & 0 & \cdots & 0 & 0 & 0 &0 &\cdots & 0& 0 \\
					I_m & 0 &  \cdots & 0 & 0 & 0 &0 & \cdots & 0& 0 \\
					0 & I_m &  \cdots & 0 & 0 & 0 &0 & \cdots & 0& 0 \\
					\vdots & \vdots & \ddots & \vdots & \vdots & \vdots & \vdots & \ddots & \vdots& \vdots \\
					0 & 0 & \cdots & I_m & 0 & 0 &0 & \cdots & 0 & 0\\
					\hline (S)_1 & (S)_2 & \cdots & (S)_{\bar{l}-1} & (S)_{\bar{l}}&  (S)_{\bar{l}+1} & (S)_{\bar{l}+2} & \cdots & (S)_{2\bar{l}-1} & (S)_{2\bar{l}} \\
					0 & 0 &  \cdots & 0 & 0 & I_p & 0 &\cdots & 0 & 0\\
					0 & 0 &  \cdots & 0 & 0 & 0& I_p &\cdots & 0 & 0\\
					\vdots & \vdots & \ddots & \vdots & \vdots& \vdots & \vdots & \ddots & \vdots &\vdots \\
					0 & 0 & \cdots & 0 & 0 &0 & 0& \cdots & I_p & 0
				\end{bmatrix}\xi_{t} + \begin{bmatrix}
					I_m \\
					0 \\
					0\\
					\vdots \\
					0\\
					\hline
					0 \\
					0 \\
					0 \\
					\vdots \\
					0
				\end{bmatrix} u_t\\
				y_t  &= S
				\xi_{t}  
			\end{aligned}
		\end{equation}
		\vspace*{4pt} 
	\end{figure*}
}

%\begin{itemize}
 %   \item $x_t$ represents the state variables which are difficult to measure directly.
%    \item $u_t$ represents the control inputs applied to the system.
 %   \item $y_t$ represents the output variables, such as active/reactive power, voltage, and current, which are easily measurable and reflect system performance.
 %   \item $w_t$ is the process disturbance affecting the system states, and $v_t$ is the measurement noise affecting the outputs.
%\end{itemize}

Our objective is to find a policy as feedback of the past input-output trajectory $u_t = \pi_t(u_{-\infty},y_{-\infty},\dots, u_{t-1}, y_{t-1})$ so that the output is regulated to zero with low control effort. 

%This is usually referred to as the stabilization problem of power converter systems in the power system control community \cite{huang2018an}.

%\begin{equation}
%\min_u \sum_{t=0}^\infty \left( \| y_t \|^2 + \| u_t \|^2 \right) dt,
%\end{equation}

\subsection{Challenges in stabilization of power converter systems (DC source + DC-AC converter + AC power grid)}

There are three main challenges in the stabilization of power converter systems:
\begin{itemize}
    \item \textbf{unknown model}: Power converter systems often exhibit high-order characteristics, rendering it difficult to establish exact dynamical models. Moreover, the power grid is also unknown when designing a controller for the converter, as the grid is large-scale and time-varying.
    
    %Specifically, the exact dynamics of the DC source and the converter are usually hard to obtain (and therefore unknown) due to proprietary manufacturer models and their complexity, e.g., the complex mechanical behaviors of wind turbines. 

    \item \textbf{measurement limitations}: Internal states are not measurable and hence we can only use input-output information for the control design.
    %\item \textbf{nonlinear interactions}: Instability can arise from nonlinear interactions between the PLL and other control loops, as well as the grid impedance, especially under weak grid conditions \cite{huang2018an}.
    \item \textbf{sudden change of dynamics}: Variations in grid conditions, operating states of power converters, and control parameter adjustments cause the power converter system to encounter sudden changes in dynamics~\cite{markovsky2023data}.
    %\item \textbf{real-Time adaptation}: The control strategy must adapt in real-time to external environmental changes, such as faults in the AC grid or varying operating conditions of power converters.
\end{itemize}

% \subsection{Control Objectives}

% The control strategy should aim to achieve the following objectives:
% \begin{itemize}
%     \item \textbf{Stabilization}: Ensure system stability by maintaining bounded state variables $x_t$, even under uncertain and varying grid conditions.
%     % \item \textbf{Output Regulation}: Minimize deviations in the output variables $y_t$ (e.g., active/reactive power, voltage, and current) from their desired reference values, while minimizing control effort $u_t$. This can be mathematically formulated as:
%     % \begin{equation}
%     % \min_u \int_0^\infty \left( \| y_t - y_{\text{ref}}_t \|^2 + \| u_t \|^2 \right) dt,
%     % \end{equation}
%     % where $y_{\text{ref}}_t$ represents the reference values for the output variables and $\| u_t \|$ is the control effort. The goal is to achieve a balance between minimizing output deviations and control input usage.
%     \item \textbf{Adaptation}: Continuously adjust control inputs $u_t$ in real-time to accommodate time-varying system dynamics and uncertainties.
%     \item \textbf{Efficiency}: Ensure the control algorithm is computationally efficient for real-time applications, avoiding delays in control actions.
% \end{itemize}

\subsection{Our approach}
In this paper, we propose a direct data-driven method to solve the stabilization problem of power converter systems without SysID. Our approach is based on data-enabled policy optimization (DeePO), an adaptive linear quadratic control method introduced in our recent work \cite{zhao2023data, zhao2024data}. It is data-driven and does not involve any explicit SysID. Moreover, DeePO uses online closed-loop data to adaptively update the control policy, making it suitable to deal with time-varying dynamics. 

However, the DeePO method \cite{zhao2024data} works only on the state-feedback setting, which is not the case for the system \eqref{equ:outputsys}. Leveraging the results from \cite{alsalti2023notes}, we first propose a controllable non-minimal realization of \eqref{equ:outputsys}, whose state is measurable from past input-output signals. Then, we propose an output-feedback DeePO algorithm for this non-minimal realization. Finally, we discuss its implementation on power converter systems and perform simulations to validate the effectiveness of the output-feedback DeePO algorithm.

\section{A controllable non-minimal realization of the output-feedback system}\label{sec: output}
In this section, we leverage the results from \cite{alsalti2023notes} to find a controllable non-minimal realization of \eqref{equ:outputsys}, where the state can be measured from input-output signals.

\subsection{A non-minimal controllable state}
We first consider the system \eqref{equ:outputsys} without noise, i.e., $d_t=0, v_t=0$. We assume that the system order $n$ and lag $l$ is unknown, but we have prior knowledge on an upper bound of the lag $\bar{l}\geq l$. Since the state is unmeasurable, we represent \eqref{equ:outputsys} with a non-minimal realization using input and output signals. Denote the input trajectory, the output trajectory, and  their stack from time $t-\bar{l}$ to $t-1$ as 
$$
{u}_{t,\bar{l}} = \begin{bmatrix}
	u_{t-1} \\ \vdots \\ u_{t-\bar{l}}
\end{bmatrix}, ~{y}_{t,\bar{l}} = \begin{bmatrix}
	y_{t-1} \\ \vdots \\ y_{t-\bar{l}}
\end{bmatrix}, ~
{\xi}_{t} = 
\begin{bmatrix}
	{u}_{t,\bar{l}} \\
	{y}_{t,\bar{l}}
\end{bmatrix},
$$
respectively. 
Define the extended observability matrix 
$$
\mathcal{O} = \begin{bmatrix}
	CA^{\bar{l}-1} \\
	\vdots \\
	CA \\
	C
\end{bmatrix},  
$$
the controllability matrix
$~\mathcal{C} = [B~~AB~~\cdots~~A^{\bar{l}-1}B]$,
and the
Toeplitz matrices capturing the impulse response
%$\mathcal{C}_w = [I~~A ~~\cdots~~A^{\bar{l}-1} ],$
%&\mathcal{T}_w = \begin{bmatrix}
%0 & C   & C A   & \cdots & C A^{\bar{l}-2}   \\
%0 & 0 & C  & \cdots & C A^{\bar{l}-3}   \\
%\vdots & \vdots & \ddots & \ddots & \vdots \\
%0 & \cdots & & 0 & C   \\
%0 & 0 & 0 & 0 & 0
%\end{bmatrix}.  
% $d_t = C(\mathcal{C}_w - A^{\bar{l}}\mathcal{O}^{\dagger}\mathcal{T}_w)w_{t,\bar{l}} -C A^{\bar{l}}\mathcal{O}^{\dagger}v_{t,\bar{l}} + v_t$
\begin{align*}
	&\mathcal{T} = \begin{bmatrix}
		0 & C B & C A B & \cdots & C A^{\bar{l}-2} B \\
		0 & 0 & C B & \cdots & C A^{\bar{l}-3} B \\
		\vdots & \vdots & \ddots & \ddots & \vdots \\
		0 & \cdots & & 0 & C B \\
		0 & 0 & 0 & 0 & 0
	\end{bmatrix}.
\end{align*}
Let $S = [C(\mathcal{C} - A^{\bar{l}}O^{\dagger}\mathcal{T}), CA^{\bar{l}}O^{\dagger}]$.
Then, we have the following results.
\begin{lemma}[\textbf{A non-minimal realization}]\label{thm:non}
	A non-minimal  realization of \eqref{equ:outputsys} is given by \eqref{equ:nonnimimal_realization}  shown at the bottom of this page.
\end{lemma}

\begin{proof}
The state can be represented with system dynamics and past input-output trajectories as
\begin{equation}\label{equ:traj}
\begin{aligned}
x_t &= A^{\bar{l}} x_{t-{\bar{l}}} + \mathcal{C}{u}_{t,{\bar{l}}}\\
{y}_{t,{\bar{l}}} &= \mathcal{O} x_{t-{\bar{l}}} + \mathcal{T}{u}_{t,{\bar{l}}}.
\end{aligned}
\end{equation}

Since ${\bar{l}}\geq l$, the extended observability matrix $\mathcal{O}$ has full column rank, and it has a unique left pseudo inverse $\mathcal{O}^{\dagger} = (\mathcal{O}^{\top}\mathcal{O})^{-1}\mathcal{O}^{\top} $. Then, it follows immediately from (\ref{equ:traj}) that  
$
x_t = (\mathcal{C} - A^{\bar{l}}\mathcal{O}^{\dagger}\mathcal{T}) {u}_{t,{\bar{l}}} + A^{\bar{l}}\mathcal{O}^{\dagger}{y}_{t,{\bar{l}}}
$
and
$
y_t = Cx_t= S\xi_{t}.
$
Thus, a non-minimal realization of \eqref{equ:outputsys} is given by \eqref{equ:nonnimimal_realization}. 
\end{proof}

However, for multiple-output systems with $p>1$, the non-minimal realization \eqref{equ:nonnimimal_realization} is generally not controllable. Moreover, the corresponding input-state data matrix 
\begin{equation}\label{equ:derank}
  \begin{bmatrix}
		u_0& u_1& \dots& u_{t-1} \\
		\xi_0& \xi_1& \dots& \xi_{t-1}
	\end{bmatrix} \in \mathbb{R}^{(m(\bar{l}+1)+p\bar{l})\times t}
\end{equation}
will never have full row rank \cite{alsalti2023notes}, which precludes the application of DeePO on the non-minimal realization \eqref{equ:nonnimimal_realization} \cite{zhao2024data}.
In fact, it can be shown that the maximal rank of \eqref{equ:derank} can only be $m(\bar{l}+1)+n$ \cite{alsalti2023notes}.

%, which contradicts the rank condition \eqref{equ:rank}. 
%
%Consequently, our previous results on the LQR problem cannot be directly adopted for the non-minimal state-feedback system \eqref{equ:nonnimimal_realization}. 

To ensure the full rank condition of data, we adopt the approach \cite{alsalti2023notes} that constructs an reduced non-minimal state for \eqref{equ:outputsys}. The following lemma is a direct implication of \cite[Theorem 4]{alsalti2023notes}.
\begin{lemma}[\textbf{A controllable non-minimal realization}]\label{lem:pert}
	Let $d_t=0, v_t=0$ for \eqref{equ:outputsys}. Then, there exists a full row rank permutation matrix $T\in \mathbb{R}^{(m\bar{l}+n) \times (m \bar{l}+p\bar{l}) }$ and $(A_z, B_z, C_z)$ such that a controllable non-minimal state-space representation is given by
		\begin{equation}\label{sys:reduce}
			\begin{aligned}
				z_{t+1} &= A_z z_t + B_z u_t \\
				y_t &= C_z z_t
			\end{aligned}
		\end{equation} 
    with a non-minimal state
        \begin{equation}\label{equ:contro_state}
			z_t = T\xi_t.
		\end{equation}
	\end{lemma}

	Since the realization \eqref{sys:reduce} is controllable, we can choose persistently exciting (PE) inputs of order $\bar{l}+1+n$ such that
	\begin{equation}\label{equ:data}
		D_{0,t} = \begin{bmatrix}
			u_0& u_1& \dots& u_{t-1} \\
			z_0& z_1& \dots& z_{t-1}
		\end{bmatrix} \in \mathbb{R}^{(m(\bar{l}+1)+n)\times t}
	\end{equation} 
	has full row rank \cite{alsalti2023notes}. Hence, we can design data-driven state-feedback controller for \eqref{sys:reduce}, where $z_t$ can be computed with the permutation $T$ and input-output signals.
      
	 Next, we show how to obtain the permutation matrix $T$ from a batch of input-output data \cite{alsalti2023notes}.
	
	\subsection{Computing the permutation matrix from input-output data}

	When there is no noise in the system \eqref{equ:outputsys},	the permutation matrix $T$ can be solved directly from a batch of input-output data \cite{alsalti2023notes}. Specifically, suppose that we have the data generated by PE input of order $\bar{l}+1+n$
		\begin{equation}\label{equ:datamat}
			\Xi_{0,t}=\begin{bmatrix}
				{\xi}_{0 } & {\xi}_{1 }  & \dots & {\xi}_{t-1 }
			\end{bmatrix} \in \mathbb{R}^{(m \bar{l}+p\bar{l}) \times t}.
		\end{equation} 
		Then, $T\in \mathbb{R}^{(m\bar{l}+n) \times (m \bar{l}+p\bar{l}) }$ is the permutation matrix such that
		$
		T \Xi_{0,t}
		$
		has full row rank $m\bar{l}+n$. When $n$ is unknown, we can easily separate linearly independent rows by, e.g., Gaussian elimination, and further obtain $T$. Moreover, the system order $n$ can be obtained as the number of rows in $T$.

	When the noise $d_t$ and $v_t$ in system \eqref{equ:outputsys} are not zero, the input-state data matrix \eqref{equ:derank} of system \eqref{equ:nonnimimal_realization} can be full row rank.
	In this case, we use singular value decomposition (SVD) for $\Xi_{0,t}$ 
	$$
	\Xi_{0,t} = \begin{bmatrix}
		U_r & U_l
	\end{bmatrix} \begin{bmatrix}
		\Lambda_r & 0 \\
		0 & \Lambda_l
	\end{bmatrix} \begin{bmatrix}
		V_r^{\top} \\ V_l^{\top}
	\end{bmatrix},
	$$
	where $\Lambda_r$ is the singular value matrix with largest $r$ singular values. 
	The number of rows $r$ of the permutation matrix $T$ should be $m\bar{l}+n$, which is consistent with   Lemma \ref{lem:pert}. Since $n$ is unknown, we can select $r > m\bar{l}$ such that there is a clear distinction between the largest $r$ singular values and the remaining ones (corresponding to noise). Then, the permutation matrix $T$ is given by
	$$
	T = \Lambda_r^{-1}U_r^{\top},
	$$
	i.e., it selects the orthogonal basis for the row space of $\Xi_{0,t}$ corresponding to largest $r$ singular values, i.e.,
	$
	V_r^{\top} = T\Xi_{0,t}.
	$

Once we obtain the permutation matrix $T$, we can measure the non-minimal state $z_t$ from input-output signals via \eqref{equ:contro_state}. Moreover, the input-state data matrix \eqref{equ:data} has full rank, which enables us to use DeePO for the realization \eqref{sys:reduce} for adaptive control.

\section{Output-feedback data-enabled policy optimization for direct adaptive control}\label{sec:deepo}
In this section, we first introduce the data-driven LQR formulation with covariance parameterization for the non-minimal realization  \eqref{sys:reduce}. Based on this, we propose a DeePO method for direct adaptive control of \eqref{equ:outputsys} using input-output data. 
  
\subsection{Direct data-driven LQR design of \eqref{sys:reduce} with covariance parameterization}
	Consider the controllable non-minimal realization \eqref{sys:reduce} with noise
\begin{equation}\label{equ:sys}
	\left\{\begin{aligned}
		z_{t+1} &= A_z z_t + B_z u_t + w_t \\
		h_t & =\begin{bmatrix}
			Q^{1 / 2} & 0 \\
			0 & R^{1 / 2}
		\end{bmatrix}
		\begin{bmatrix}
			z_t \\
			u_t
		\end{bmatrix}
	\end{aligned}\right. 
\end{equation}
Here, $h_t$ is the performance signal of interest  and the weighting matrices $(Q, R)$ are positive definite. 

The LQR problem is phrased as finding a state-feedback gain $K\in \mathbb{R}^{m\times (m\bar{l}+n)}$ that minimizes the $\mathcal{H}_2$-norm of the transfer function $\mathscr{T}(K):w \rightarrow h$ of the closed-loop system
$$
\begin{bmatrix}
	z_{t+1} \\
	h_t
\end{bmatrix}=\begin{bmatrix}
	A_z+B_zK & I_n \\
	\hline \begin{bmatrix}
		Q^{1 / 2} \\
		R^{1 / 2} K
	\end{bmatrix} & 0
\end{bmatrix}\begin{bmatrix}
	z_t \\
	d_t
\end{bmatrix}.
$$
When $A_z+B_zK$ is stable, it holds that \cite{anderson2007optimal}
\begin{equation}\label{equ:transfer}
	\|\mathscr{T}(K)\|_2^2  = \text{Tr}((Q+K^{\top}RK)\Sigma_K)=:J(K),
\end{equation}
where $\Sigma_K$ is the closed-loop state covariance matrix obtained as the positive definite solution to the Lyapunov equation
\begin{equation}\label{equ:Sigma}
	\Sigma_K = I_n + (A_z+B_zK)\Sigma_K (A_z+B_zK)^{\top}.
\end{equation}
We refer to $J(K)$ as the LQR cost and to (\ref{equ:transfer})-(\ref{equ:Sigma}) as a \textit{policy parameterization} of the LQR. 
When the model parameters are known, the optimal LQR gain $K^*$ is unique and can be found by, e.g., solving an algebraic Riccati equation~\cite{anderson2007optimal}. 

Since $(A_z, B_z)$ is unknown, data-driven methods learn the LQR gain from input-state data. Consider the $t$-long time series of states, inputs, noises, and successor states of \eqref{sys:reduce}
\begin{align*}
	Z_{0,t} &:= \begin{bmatrix}
		z_0& z_1& \dots& z_{t-1}
	\end{bmatrix}\in \mathbb{R}^{(m\bar{l}+n)\times t},\\
	U_{0,t} &:= \begin{bmatrix}
		u_0& u_1& \dots& u_{t-1}
	\end{bmatrix}\in \mathbb{R}^{m\times t}, \\
	W_{0,t} &:= \begin{bmatrix}
		w_0& w_1& \dots& w_{t-1}
	\end{bmatrix}\in \mathbb{R}^{(m\bar{l}+n)\times t}, \\
	Z_{1,t} &:= \begin{bmatrix}
		 z_1& z_2& \dots& z_t
	\end{bmatrix}\in \mathbb{R}^{(m\bar{l}+n)\times t},
\end{align*}
which satisfy the system dynamics
\begin{equation}\label{equ:dynamics}
	Z_{1,t} = A_zZ_{0,t}+ B_zU_{0,t} + W_{0,t}.
\end{equation}

Assume that the input data is PE of order $\bar{l}+1+n$. Then, the block matrix
$
[U_{0,t}^{\top},Z_{0,t}^{\top}]^{\top} 
$
has full row rank. Define the sample covariance of input-state data as
\begin{equation}
	\Phi_t:=\frac{1}{t}\begin{bmatrix}
		U_{0,t} \\ Z_{0,t}
	\end{bmatrix}\begin{bmatrix}
		U_{0,t} \\ Z_{0,t}
	\end{bmatrix}^{\top},
\end{equation}
which is positive definite due to the full rank condition. Then, we can use sample covariance to parameterize the policy 
\begin{equation}\label{equ:newpara}
\begin{bmatrix}
	K \\
	I_{(m\bar{l}+n)}
\end{bmatrix}=    \Phi_t V,
\end{equation}
where $V\in \mathbb{R}^{(m\bar{l}+m+n)\times (m\bar{l}+m+n)}$.

With the covariance parameterization \eqref{equ:newpara}, the LQR problem (\ref{equ:transfer})-(\ref{equ:Sigma}) can be expressed by raw data matrices $(Z_{0,t}, U_{0,t}, Z_{1,t})$ and the optimization matrix $V$. For brevity, let $\overline{Z}_{0,t}= Z_{0,t}D_{0,t}^{\top}/t$ and $\overline{U}_{0,t}=  U_{0,t}D_{0,t}^{\top}/t$ be a partition of $\Phi_t$, and let
$\overline{W}_{0,t}=  W_{0,t}D_{0,t}^{\top}/t$ be the noise-state-input covariance, and finally define the covariance with respect to the successor state $\overline{Z}_{1,t}=  Z_{1,t}D_{0,t}^{\top}/t$.  
Then, the closed-loop matrix can be written as
$$
[B_z,A_z]\begin{bmatrix}
	K \\
	I_{m\bar{l}+n}
\end{bmatrix}\overset{\eqref{equ:newpara}}{=}[B_z,A_z]\Phi_t V\overset{\eqref{equ:dynamics}}{=}(\overline{X}_{1,t} - \overline{W}_{0,t})V.
$$
Following the certainty-equivalence principle~\cite{dorfler2021certainty}, we disregard the
unmeasurable  $\overline{W}_{0,t}$ and use $\overline{Z}_{1,t}V$ as the closed-loop matrix. After substituting $A+BK$  with $\overline{Z}_{1,t}V$ in (\ref{equ:transfer})-(\ref{equ:Sigma}) and leveraging \eqref{equ:newpara}, the LQR problem becomes 
\begin{equation}\label{prob:equiV}
	\begin{aligned}
		&\mathop{\text {minimize}}\limits_{V}~J_t(V) :=\text{Tr}\left((Q+V^{\top}\overline{U}_{0,t}^{\top}R\overline{U}_{0,t}V)\Sigma_t\right),\\
		&\text{subject to}~ ~\overline{Z}_{0,t}V= I_{m\bar{l}+n},
	\end{aligned}
\end{equation}
where $\Sigma_t = I_{m\bar{l}+n} + \overline{Z}_{1,t}V\Sigma_t V^{\top}\overline{Z}_{1,t}^{\top}$ is a covariance parameterization of \eqref{equ:Sigma},
and the gain matrix can be recovered by $K = \overline{U}_{0,t}V$. We refer to (\ref{prob:equiV}) as the direct data-driven LQR problem, which does not involve any explicit SysID.

\subsection{Output-feedback DeePO}
Policy optimization (PO) refers to a class of direct design methods, where the policy is parameterized and recursively updated using gradient methods~\cite{bin2022towards}. In particular, the DeePO algorithm uses online gradient descent of \eqref{prob:equiV} to recursively update $V$. The details are presented in Algorithm \ref{alg:inputoutput}. Given offline data, we first compute the permutation matrix $T$. At time $t$, we apply the linear state feedback policy $u_t=K_tz_t + e_t$ for control, where $e_t\in \mathbb{R}^m$ is a probing noise used to ensure the PE rank condition. We use online projected gradient descent to update the parameterized policy, where the projection  $\Pi_{\overline{Z}_{0,t+1}}: = I_{m(\bar{l}+1)+n}-\overline{Z}_{0,t+1}^{\dagger}\overline{Z}_{0,t+1}$ onto the nullspace of $\overline{Z}_{0,t+1}$ is to ensure the subspace constraint in \eqref{prob:equiV}.
 Define the feasible set of \eqref{prob:equiV} (i.e., the set with stable closed-loop matrices) as $\mathcal{S}_t:= \{V\mid \overline{Z}_{0,t}V =I_{m\bar{l}+n},  \rho (\overline{Z}_{1,t}V)<1\}$. The gradient can be computed by the following lemma.
\begin{lemma}[{\cite[Lemma 2]{zhao2024data}}]\label{lem:gradient}
	For $V\in \mathcal{S}_t$, the gradient of $J_t(V)$ with respect to $V$ is given by
	\begin{equation}\label{equ:pg}
		\nabla J_t(V) = 2 \left(\overline{U}_{0,t}^{\top}R\overline{U}_{0,t}+\overline{Z}_{1,t}^{\top}P_t\overline{Z}_{1,t}\right)V \Sigma_t,
	\end{equation}
	where $P_t$ satisfies the Lyapunov equation 
	$$
		P_t = Q + V^{\top}\overline{U}_{0,t}^{\top}R\overline{U}_{0,t}V + V^{\top}\overline{Z}_{1,t}^{\top}P_t\overline{Z}_{1,t}V.
    $$
\end{lemma}

By Lemma \ref{lem:gradient}, the gradient can be computed by solving two Lyapunov equations. The stepsize \(\eta_t\) should be chosen based on the signal-to-noise ratio (SNR) of the online data. Specifically, when the SNR is high, the gradient direction is more reliable, allowing for a larger stepsize. Conversely, when the SNR is low, a smaller stepsize is necessary to prevent the policy from deviating from the stability region. Accordingly, we set the stepsize as
	\begin{equation}\label{equ:stepsize} 
		\eta_t = \frac{\eta_0}{\left\|\overline{U}_{0,t}\Pi_{\overline{Z}_{0,t}}\overline{U}_{0,t}^{\top}\right\|}, ~t\geq t_0,
	\end{equation}
	where $\eta_0$ is a constant, and the denominator is used to quantify the SNR.

The DeePO algorithm is  direct and adaptive  in the sense that it directly uses online closed-loop data to update the policy. Thus, it can rapidly adapt to dynamic changes reflected in the data. Algorithm \ref{alg:inputoutput} has a recursive policy update and can be implemented efficiently. Specifically, all covariance matrices and the inverse $\Phi_{t+1}^{-1}$ have recursive updates. Moreover, the parameterization can be updated recursively via rank-one update, i.e.,
$$
V_{t+1}= \frac{t+1}{t}\left(V_{t}' - \frac{\Phi_{t}^{-1}\phi_t\phi_t^{\top}V_{t}' }{t+\phi_t^{\top}\Phi_{t}^{-1}\phi_t}\right),
$$
where $\phi_t = [u_{t}^{\top}, z_{t}^{\top}]^{\top}$, and $\Phi_{t}^{-1}$ and $V_{t}'$ are given from the last iteration. Theoretically, it is shown that under mild assumptions the policy $\{K_t\}$ converges to the optimal LQR gain. We refer to \cite[Section IV]{zhao2024data} for detailed discussions.
  
\begin{algorithm}[t]
	\caption{Output-feedback DeePO}
	\label{alg:inputoutput}
	\begin{algorithmic}[1]
		\Require Offline   data $(u_{-\bar{l}}, y_{-\bar{l}}, \dots, u_{t_0-1}, y_{t_0-1}, y_{t_0})$ and a stepsize $\eta_t$.
		\State \textbf{Computation of the permutation matrix:} Constitute the data matrix $\Xi_{0,t_0}$ in \eqref{equ:datamat} and perform SVD
		$$
		\Xi_{0,t_0} = \begin{bmatrix}
			U_r & U_l
		\end{bmatrix} \begin{bmatrix}
			\Lambda_r & 0 \\
			0 & \Lambda_l
		\end{bmatrix} \begin{bmatrix}
			V_r^{\top} \\ V_l^{\top}
		\end{bmatrix},
		$$
		where $r > m\bar{l}$ is selected such that there is a clear distinction between the largest $r$ singular values and the remaining ones. Let $
		T = \Lambda_r^{-1}U_r^{\top}.
		$ 
        \State Compute the initial policy $K_{t_0}$ from \eqref{prob:equiV} with offline data.
		\For{$t=t_0,t_0+1,\dots$}
		\State Compute $z_t = T \xi_t$, apply $u_t=K_tz_t + e_t$ to the system, and observe $z_{t+1}$.
		\State \textbf{Policy parameterization:} given $K_{t}$, solve $V_{t+1}$ via 
		$$
		V_{t+1} =\Phi_{t+1}^{-1} \begin{bmatrix}
			K_{t} \\
			I_n
		\end{bmatrix}.
		$$
		\State \textbf{Update of the parameterized policy:} perform one-step projected gradient descent
		$$
		V_{t+1}' = V_{t+1} - \eta_t  \Pi_{\overline{Z}_{0,t+1}} \nabla J_{t+1}(V_{t+1}).
		$$ 
		\State \textbf{Gain update:} update the control gain by 
		$$
		K_{t+1} = \overline{U}_{0,t+1}V_{t+1}'.
		$$
		\EndFor	
	\end{algorithmic}
\end{algorithm}

\section{DeePO for stabilization of power converters and renewable energy systems}\label{sec: sim}
In this section, we first discuss the implementation of output-feedback DeePO on the power converter systems. Then, we perform simulations for a grid-connected power converter and a direct-drive wind generator.
\subsection{Implementation of output-feedback DeePO for power converter systems}

We consider two typical power converter systems: a grid-connected power converter (with lithium batteries as the DC source) in Fig. \ref{fig.1} and a direct-drive (type 4) wind generator in Fig. \ref{fig.3}, which also uses a converter as the grid interface. The maximum lags for the two systems are $\bar{l} = 2$ and $4$, respectively. Both systems share similar control challenges, including unknown models, measurement limitations, and potential change of dynamics. While Algorithm~\ref{alg:inputoutput} considers only time-invariant systems, it can potentially handle changes in the system dynamics. This is because DeePO uses online closed-loop data to update the policy in real time and can quickly adapt to changes. Moreover, since Algorithm~\ref{alg:inputoutput} has a recursive implementation, the online update of the policy is computationally efficient.

For both systems, the outputs of the unknown dynamics are defined as the deviations of active and reactive power from their reference values: $y = [P_E - P_0, Q_E - Q_0]^\top$. The DeePO method is configured to provide two control
inputs added to the current references of the system. All input and output signals are expressed in per-unit ($\mathrm{p.u.}$) values. Since internal states are not directly measurable, the input-output stack is constructed as $\xi_t = [u_{t,\bar{l}}^\top, y_{t,\bar{l}}^\top]^\top$. The matrix $\Xi_{0,t}$ is built from past trajectories, followed by SVD to determine the reduced dimension $r$. A projection matrix $T$ is then computed to obtain the reduced-order state $z_t$ from $\xi_t$.

The parameters for the two systems are set as follows:
\begin{itemize}
    \item  The sampling frequency is chosen as $200\,\mathrm{Hz}$.
    \item  The trajectory length is set to $300$ for the power converter and $600$ for the wind generator.
    \item  The input and state weighting matrices are set as $R = I_m$ and $Q = I_r$, respectively.
    \item  The stepsize for gradient descent is set according to \eqref{equ:stepsize} with $\eta_0 = 0.0001$.  
     %\item \textbf{Control Horizon}: The control horizon is set as $k = 1$ for both systems, meaning the computed inputs are applied to the system before solving DeePO again.
\end{itemize}

% Both applications benefit from the adaptive nature of DeePO, which uses online closed-loop data to update the control policy in real time. 

% This provides a computationally efficient mechanism for dealing with time-varying system dynamics, and ensures the algorithm's flexibility and responsiveness to changes in grid and system conditions. 

% With this unified DeePO configuration, we proceed to explore the specific implementation details and performance results of each system.

\subsection{Simulations on a grid-connected power converter}

The grid-connected power converter system considered in this study is shown in Fig. \ref{fig.1}, where we employ the DeePO controller. The objective of the DeePO controller is to regulate the deviations of active and reactive power from their references to zero with minimal control effort, thereby effectively mitigating power oscillations caused by grid disturbances.

%The system has a total of $11$ state variables, the majority of which cannot be directly measured. Therefore, input-output information is used to construct a non-minimal realization of the system. New state variables are constructed as $z_{t} = [u_{t,T_{\mathrm{ini}}}^{\top},y_{t,T_{\mathrm{ini}}}^{\top}]^{\top} $, where $T_{\mathrm{ini}}=6$. Although $T_{\mathrm{ini}}$ is smaller than the theoretical order of the system, the dynamics of the dominant states for the studied problem (oscillation damping) is still able to be captured \cite{markovsky2023data}, which will be further verified in subsequent tests.

\begin{figure}[t] 
    \centering 
    \includegraphics[width=0.95\linewidth]{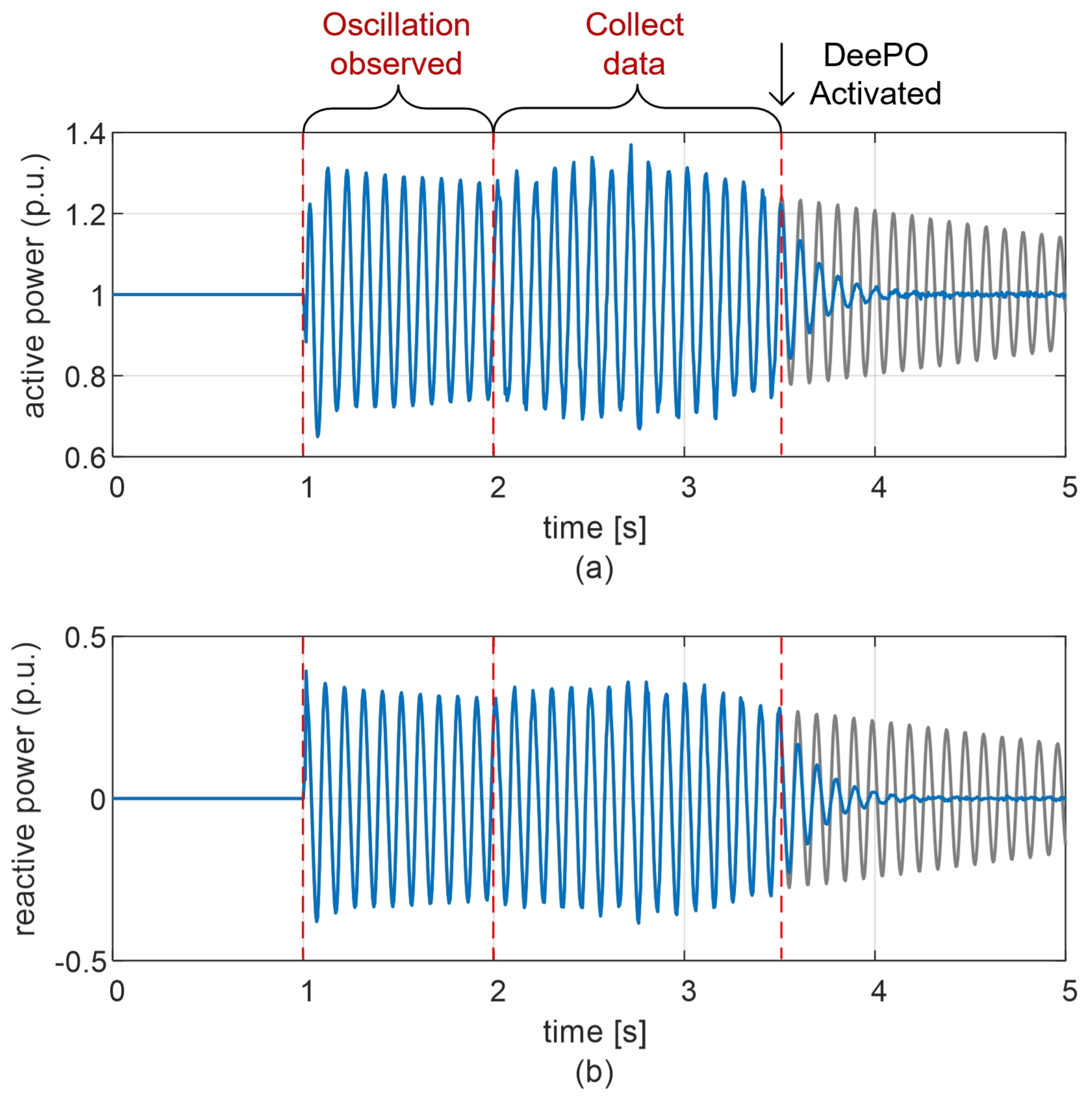} 
    \vspace{-3mm}
    \caption{Time-domain responses of the grid-connected power converter. The DeePO is activated at $t = 3.5\,\mathrm{s}$.
    \textcolor[HTML]{0070C0}{\textbf{---}} with DeePO;
    \textcolor[HTML]{7F7F7F}{\textbf{---}} without DeePO.} 
    \label{fig.2} 
\end{figure}

Fig. \ref{fig.2} shows the time-domain responses of the power converter. At $t = 1.0~\mathrm{s}$, we change the short circuit ratio of the system from $5$ to $2.21$ to emulate an event in the power grid, for example, tripping of transmission lines. It can be seen that the converter starts to oscillate after the disturbance, which is caused by the interactions among PLL, current/power control loops, and the weak power grid~\cite{huang2018an}. Such oscillations are undesirable, as they may trigger resonance in the system, increase the risk of equipment failure, and even result in grid collapse.

After the oscillation is observed, we inject band-limited white noise signals into the system through the two input channels from $t = 2.0\,\mathrm{s}$ to $t = 3.5\,\mathrm{s}$ to excite the system and collect data. We then construct the Hankel matrix $\Xi_{0,t}$ using the input-output data from this interval and perform SVD. The resulting singular values, in descending order, are~$7.075,\ 2.596,\ 0.556,\ 0.496,\ 0.476,\ 0.460,\ 0.249,\ 0.186$. The minimum embedding dimension is $m\bar{l} = 4$. Since the first six singular values exhibit a clear distinction from the remaining two, we select $r = 6$. The projection matrix $T$ is then computed based on the leading components, and the reduced-order state is obtained accordingly. This representation is used to solve an SDP problem of \eqref{prob:equiV}, yielding $K_{t_0}$ as the initial policy for DeePO. The DeePO controller is activated at $t = 3.5\,\mathrm{s}$, providing real-time control inputs that effectively counteract the oscillation. As depicted in Fig. \ref{fig.2}, DeePO effectively damps the oscillation and restores the stable operation of the power converter. In contrast, the responses of the power converter without DeePO are shown as the grey lines in Fig. \ref{fig.2}, where the system exhibits sustained oscillations, indicating a low stability margin and increasing the risk of equipment damage and potential cascading failures in the power system.

%It is worth noting that after the activation of DeePO, the noise signals can be reduced but are still required to collect data for continuous control strategy updates.

\subsection{Simulations on a direct-drive wind generator}
\begin{figure}[t] 
    \centering 
    \includegraphics[width=0.98\linewidth]{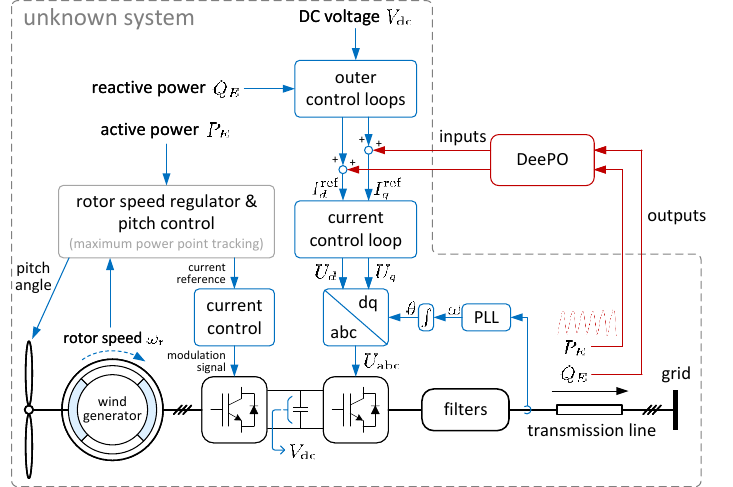} 
    \vspace{-3mm}
    \caption{The application of DeePO to stabilize a direct-drive wind generator.}
    \label{fig.3} 
\end{figure}

We consider now a direct-drive wind generator with a high-fidelity model implemented in MATLAB/Simulink (2023b)~\cite{markovsky2023data}. The wind generator model here is more complicated than the converter model in the last subsection, as it not only has a converter as the grid interface, but also considers the generator dynamics on the DC side.
The simulation model includes detailed turbine dynamics, flux dynamics, filters, speed control, pitch control, converter control, and maximum power point tracking (MPPT). The grid is modeled as a voltage source behind a transmission line with unknown impedance, which impacts the wind generator in a closed-loop manner. The whole system is illustrated in Fig. \ref{fig.3}. Due to proprietary manufacturer models and the complexity of the power grid, its exact dynamical model can hardly be derived or identified. 
In what follows, we apply the DeePO method in Algorithm \ref{alg:inputoutput} to stabilize the wind generator. 

\begin{figure}[t] 
    \centering 
    \includegraphics[width=0.95\linewidth]{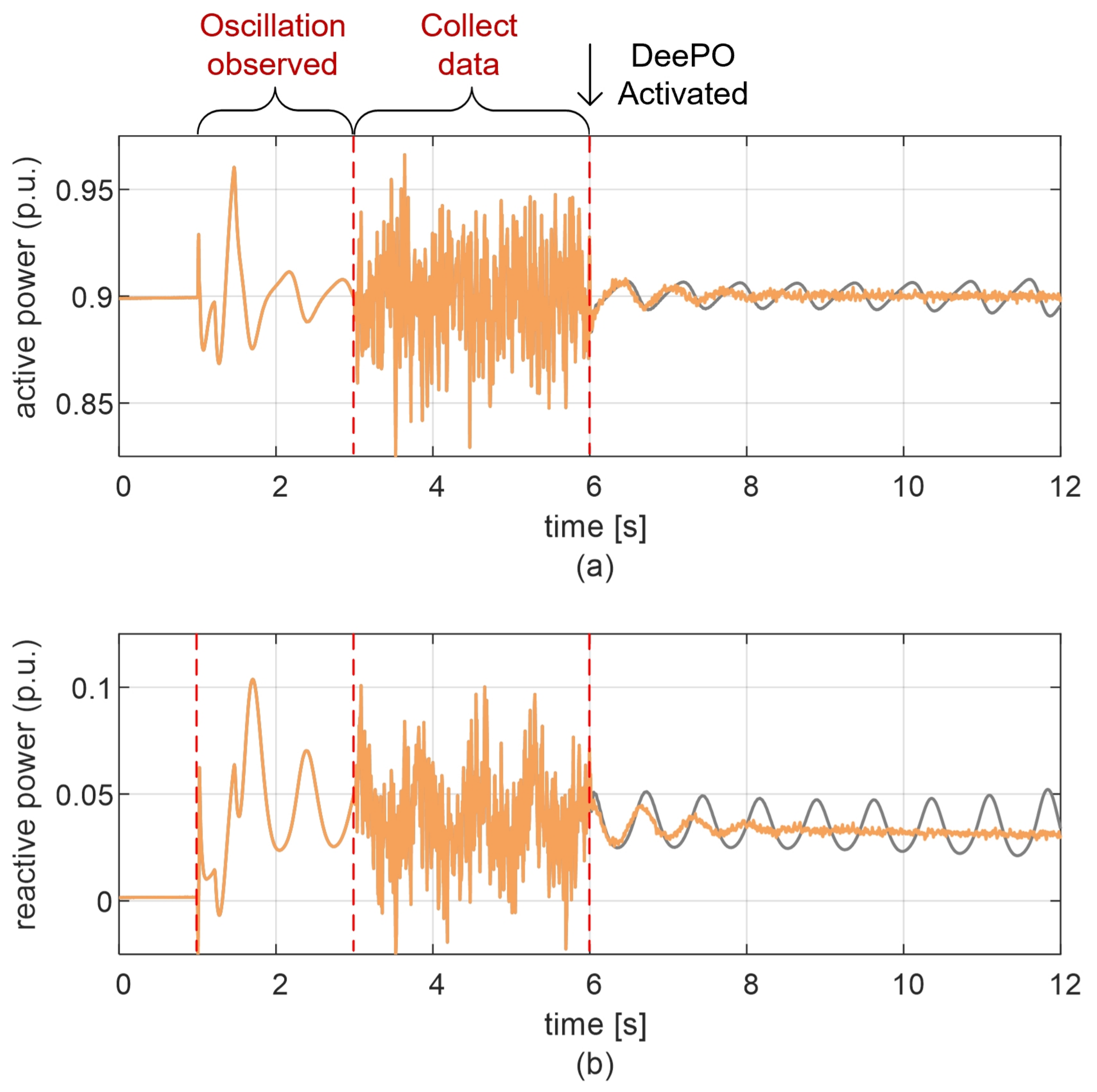}
    \vspace{-3mm}
    \caption{The time-domain responses of the wind generator: (a) active power and (b) reactive power. The DeePO is activated at $t = 6.0\,\mathrm{s}$.
    \textcolor[HTML]{F7A25B}{\textbf{---}} with DeePO;
    \textcolor[HTML]{7F7F7F}{\textbf{---}} without DeePO.} 
    \label{fig.4} 
\end{figure}

Fig. \ref{fig.4} shows the time-domain responses of the wind generator under different configurations of DeePO. At $t = 1.0\,\mathrm{s}$, the short circuit ratio is changed from $2.23$ to $2$, simulating a grid event, such as the tripping of transmission lines. Following the disturbance, the wind generator begins to oscillate. 

Then, we inject band-limited white noise to collect input-output trajectory data from $t = 3.0\,\mathrm{s}$ to $t = 6.0\,\mathrm{s}$. Following the same procedure as in the last subsection, we determine the embedding dimension as $r = 12$ for this case. Based on the data collected in this interval, we solve an SDP of \eqref{prob:equiV} to obtain the initial DeePO policy $K_{t_0}$. As shown in Fig.~\ref{fig.4},  DeePO is activated at $t = 6.0\,\mathrm{s}$ and successfully provides real-time control inputs, eliminating the oscillations. By comparison, the wind generator system exhibits a low stability margin without DeePO, with persistent oscillations in active and reactive power after $t = 6.0\,\mathrm{s}$. These sustained oscillations not only threaten the integrity of the wind generator itself, due to potential mechanical resonance, but also compromise the secure operation of the overall power system. We observe that the damping performance of DeePO is comparable to that of DeePC (c.f. Fig.~14 in~\cite{markovsky2023data}). Nevertheless, DeePO offers a significant computational advantage. For instance, on an AMD Ryzen 7 PRO 7840H CPU with 32~GB of RAM, solving a quadratic program online (required by DeePC) takes approximately $15~\mathrm{ms}$ per time step, whereas DeePO only requires around $1~\mathrm{ms}$ to perform a gradient step. This highlights the efficiency of DeePO in delivering similar control performance at a much lower computational cost.

\begin{figure}[t] 
    \centering 
    \includegraphics[width=0.95\linewidth]{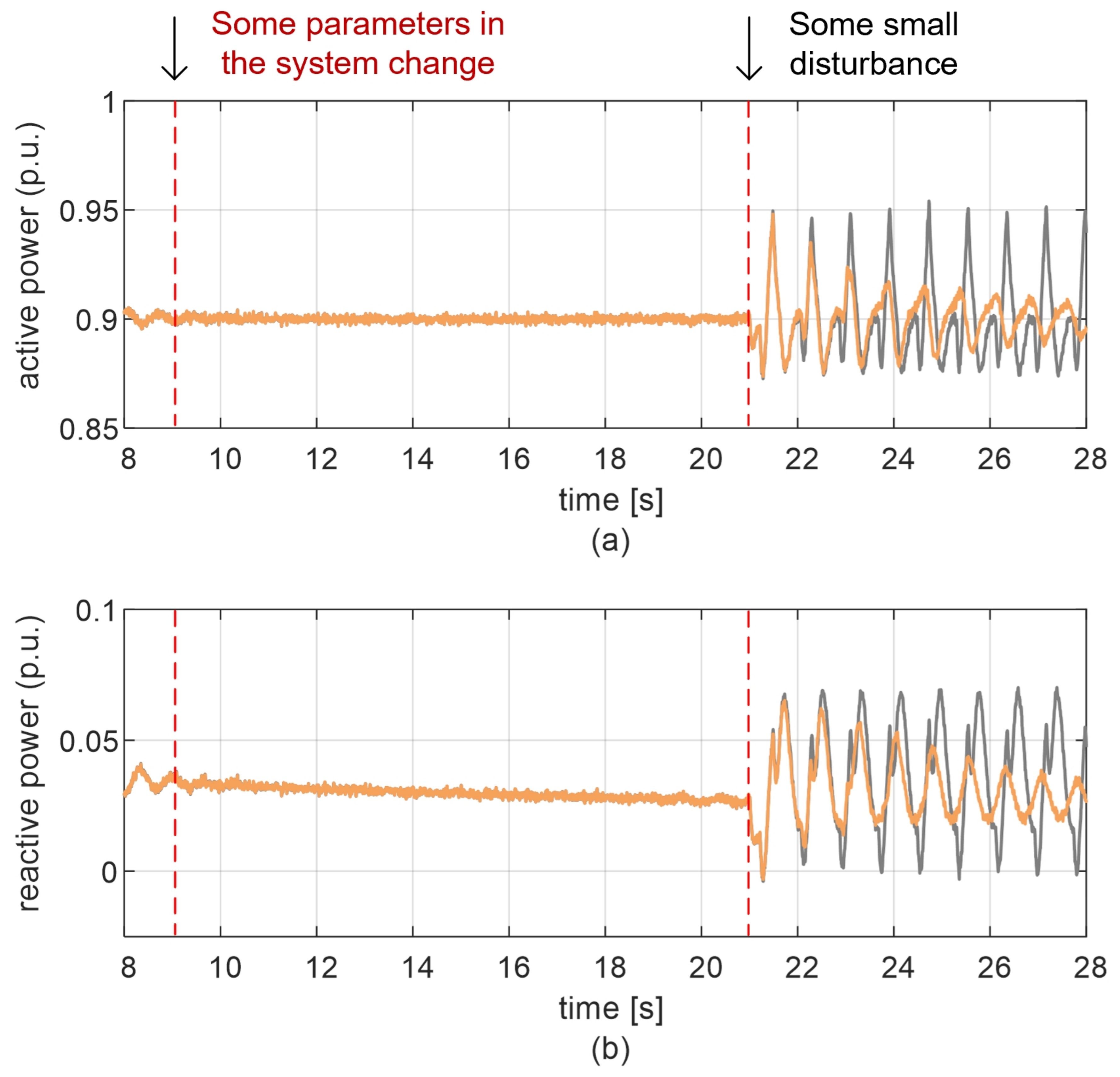} 
    \vspace{-3mm}
    \caption{The time-domain responses of the wind generator: (a) active power and (b) reactive power. 
    \textcolor[HTML]{F7A25B}{\textbf{---}} with DeePO controller;
    \textcolor[HTML]{7F7F7F}{\textbf{---}} with non-adaptive controller.} 
    \label{fig.5} 
\end{figure}

In time-varying environments, non-adaptive controllers that rely solely on historical data may fail to maintain effective control. To illustrate this, we compare a non-adaptive controller with DeePO under changing system conditions. Both controllers apply the same initial policy $K_{t_0}$ up to $t = 9.0\,\mathrm{s}$. At this point, the DC voltage control parameters are reduced to 90\% of their nominal values, altering the system dynamics without affecting the equilibrium point. After $t = 9.0\,\mathrm{s}$, the non-adaptive controller retains the fixed gain $K_{t_0}$, while DeePO updates its policy online using real-time input-output data. Fig.~\ref{fig.5} shows the time-domain responses under this time-varying condition and a subsequent disturbance introduced at $t = 21.0\,\mathrm{s}$ by injecting a \textit{q}-axis voltage perturbation into the PLL. While the non-adaptive controller exhibits degraded performance, DeePO successfully maintains stability and quickly damps out oscillations. This improvement stems from its ability to continuously adapt to changing dynamics in real time.

%\subsection{Simulation on a direct-drive wind generator}
%Next, we apply the DeePO algorithm to stabilize a direct-drive (type 4) wind generator. The conventional control design for wind generators is typically model-based, assuming an ideal power grid with fixed voltage magnitude and frequency. However, such controllers often perform poorly or may become unstable when connected to non-ideal grids. While robust and adaptive model-based methods exist to handle these issues, they tend to result in complex controllers and may still struggle to adapt to all grid conditions.

%In this case, we adjust the transmission line length to simulate a weak power grid, with a short circuit ratio of 2.23. The wind generator relies on a synchronous reference frame PLL to detect the voltage phase and achieve synchronization. To avoid strongly time-varying dynamics, we disable the automatic gain control and fix the PLL bandwidth to 60 rad/s. Under the weak grid condition, the wind generator exhibits low stability and may become unstable when the power grid changes.

%In such cases, where the system parameters evolve, non-adaptive control may struggle to maintain performance across different operating conditions. To address this, we now demonstrate the advantages of the adaptive DeePO, which updates its control policies recursively in real time. The following tests will highlight how adaptive DeePO outperforms the non-adaptive variant by better accommodating time-varying dynamics.

\section{Conclusion}\label{sec:conclu}
In this paper, we proposed a output-feedback data-enabled policy optimization (DeePO) method, which is direct data-driven, adaptive, based on online input-output data, and recursive in terms of implementation. We demonstrated via simulations the effectiveness of DeePO in stabilizing and mitigating undesired oscillations in a grid-connected power converter and a direct-drive wind generator. 

%In particular, we observed that DeePO has satisfactory performance in stabilization, and shows a similar damping ratio to data-enabled predictive control (DeePC). Nevertheless, DeePO is computationally more efficient than DeePC, as it is gradient-based and does not require solving optimization problems online.

Future work includes the extension of the output-feedback DeePO to achieve other control objectives (e.g., reference tracking) and to other types of systems (e.g., slowly time-varying systems). It is also valuable to investigate the effect of regularization on output-feedback DeePO of power converter systems \cite{zhao2025regularization}.

\bibliographystyle{IEEEtran}
\bibliography{mybibfile}

\end{document}